\PassOptionsToPackage{table,xcdraw}{xcolor}

\documentclass[sigconf]{acmart}

\pdfoutput=1

\AtBeginDocument{%
  \providecommand\BibTeX{{%
    \normalfont B\kern-0.5em{\scshape i\kern-0.25em b}\kern-0.8em\TeX}}}

\copyrightyear{2024}
\acmYear{2024}
\setcopyright{acmlicensed}\acmConference[CHI '24]{Proceedings of the CHI Conference on Human Factors in Computing Systems}{May 11--16, 2024}{Honolulu, HI, USA}
\acmBooktitle{Proceedings of the CHI Conference on Human Factors in Computing Systems (CHI '24), May 11--16, 2024, Honolulu, HI, USA}
\acmDOI{10.1145/3613904.3642238}
\acmISBN{979-8-4007-0330-0/24/05}

\acmSubmissionID{1971}

\usepackage{appendix}

\usepackage{subfigure}

\usepackage{graphicx}
\usepackage{multirow}
\usepackage{pifont}
\usepackage{lscape}

\usepackage{mathtools}

\begin{document}

\title[“It’s Kind of Context Dependent”: Understanding BLV People’s Video Accessibility Preferences Across Viewing Scenarios]{“It’s Kind of Context Dependent”: Understanding Blind and \\ Low Vision People’s Video Accessibility Preferences Across Viewing Scenarios}

\author{Lucy Jiang}
\affiliation{%
  \institution{Cornell University}
  \city{Ithaca, NY}
  \country{USA}}
\email{lucjia@cs.cornell.edu}

\author{Crescentia Jung}
\affiliation{%
  \institution{Cornell Tech}
  \city{New York, NY}
  \country{USA}}
\email{cj382@cornell.edu}

\author{Mahika Phutane}
\affiliation{%
  \institution{Cornell University}
  \city{Ithaca, NY}
  \country{USA}}
\email{mahika@cs.cornell.edu}

\author{Abigale Stangl}
\affiliation{%
  \institution{Georgia Institute of Technology}
  \city{Atlanta, GA}
  \country{USA}}
\email{abigale.stangl@design.gatech.edu}

\author{Shiri Azenkot}
\affiliation{%
  \institution{Cornell Tech}
  \city{New York, NY}
  \country{USA}}
\email{shiri.azenkot@cornell.edu}

\renewcommand{\shortauthors}{Jiang et al.}

\begin{abstract}
While audio description (AD) is the standard approach for making videos accessible to blind and low vision (BLV) people, existing AD guidelines do not consider BLV users’ varied preferences across viewing scenarios. These scenarios range from how-to videos on YouTube, where users seek to learn new skills, to historical dramas on Netflix, where a user’s goal is entertainment. Additionally, the increase in video watching on mobile devices provides an opportunity to integrate nonverbal output modalities (e.g., audio cues, tactile elements, and visual enhancements). Through a formative survey and 15 semi-structured interviews, we identified BLV people’s video accessibility preferences across diverse scenarios. For example, participants valued action and equipment details for how-to videos, tactile graphics for learning scenarios, and 3D models for fantastical content. We define a six-dimensional video accessibility design space to guide future innovation and discuss how to move from “one-size-fits-all” paradigms to scenario-specific approaches.
\end{abstract}




\maketitle
\section{Introduction}
As videos become more widespread and stylistically diverse, ranging from documentaries on Netflix to 60-second videos on Instagram, blind and low vision (BLV) people remain excluded from engaging with this growing variety and volume of visual content. Currently, the standard method of making videos accessible is adding audio description (AD), a separate audio track with narration of visual elements \cite{acbadp, snyder2005audio, fryer2016introduction}. AD is typically created for high-budget content such as movies and TV shows. However, many videos still lack AD, and it is unclear to what extent existing accessibility practices can or should be applied to newer video formats such as short-form video. As user-generated content increases in popularity on platforms like YouTube and TikTok \cite{ytstats, ytbillion, forbesvideo}, the need to make videos of all types accessible to BLV people grows. 

To support video accessibility, researchers and practitioners have aimed to increase the \textit{quantity} of described videos through crowdsourcing platforms and automation (e.g., \cite{youdescribe, natalie2021efficacy, bodi2021automated, campos2020cinead, wang2021toward, killough2023exploring}). Others have focused on improving the \textit{quality} of AD through providing authorship guidelines to help determine what content to include or which tone of voice to use (e.g., \cite{dcmpguidelines, descriptionkeydcmp, adcoalition, acbadpguide, dcmphowtodescribe, jiang2023beyond, jiang2022co}). However, limited research has explored how nonverbal techniques, which include visual, audio, and tactile enhancements \cite{sackl2020ensuring, jiang2023beyond}, can support video accessibility in a holistic manner.

Furthermore, no work has systematically considered the evolving ways in which people consume videos today. Over the last two decades, video consumption has shifted from only watching on large screens to frequent watching on mobile devices \cite{nevsky2023accessibility, rigby2016watching}. Currently, people often watch videos on mobile devices or computers and use platforms that allow them to access new content at an unprecedented rate. Video types have also become more diverse \textemdash{} videos can range from 5-minute how-to videos on YouTube, to hour-long documentaries on Netflix, to comedic 30-second short videos on TikTok. Users’ goals for watching these different videos include learning how to do something, being entertained, or keeping up with family and friends. 

We conceptualize these video watching contexts as \textbf{\textit{viewing scenarios}}, which encapsulate (1) video types (e.g., how-to, comedy, music video), (2) viewing platforms (e.g., streaming services, video sharing sites, social networking sites), and (3) users’ information goals (e.g., to learn, to be entertained, to engage with friends). In other words, a scenario is the story of what video is being watched, where a user found the video, and why a user is watching the video. Researchers have established that BLV people’s accessibility needs vary for different image viewing scenarios \cite{stangl2021going}; however, few have explored the nuances of accessibility for the large variety of videos produced today. Thus, there is a gap in understanding how BLV people’s video accessibility needs vary across different scenarios. 

To address this gap, we investigate the following research question: \textbf{What are BLV people’s needs and preferences for video accessibility across viewing scenarios?} We considered various approaches to video accessibility, including the content and presentation of AD and augmented output modalities, to build on prior work \cite{jiang2023beyond}. We conducted a formative survey with 101 respondents and interviewed 15 BLV participants. Interviews included a discussion of recently watched videos and a co-watching session encompassing multiple video scenarios. Throughout the interview, we probed about current access needs and brainstormed ideas for holistically enhancing video accessibility. 

We found that BLV user needs and preferences for video accessibility varied across viewing scenarios. For example, participants wished to know details about actions and equipment to help with learning how to do something from how-to videos on YouTube. In contrast, when watching short-form videos on Instagram or Facebook to engage with friends, participants placed more emphasis on subjects, actions, clothing, and settings. Participants’ desired output modalities ranged from standardizing audio cues for indicating scene changes to using physical and tangible 3D models for conceptualizing fantastical character designs. Across scenarios, we also identified that video types strongly correlated with platforms used and users’ goals. Based on our findings, we present six dimensions in the video accessibility design space: level of detail, alteration of video time, level of augmentation, modality of presentation, synchronicity of accessible content, and tone and style of approach. We also consider the efficacy and ethical implications of using generative AI to support video accessibility for a subset of dimensions in our design space.

In summary, we (1) contribute novel insights on the variety and specificity of BLV users’ preferences for video accessibility across diverse \textit{viewing scenarios}, (2) articulate a six-dimensional design space for holistic video accessibility, and (3) consider how advancements in AI technology intersect with video and content accessibility efforts. Our work challenges the existing paradigm of “one-size-fits-all” audio descriptions. We intend for our design space to serve as a valuable resource for the accessibility community as videos, media, and technology continue to evolve. 

\section{Related Work}
Our work builds on prior work in image and video accessibility, specifically regarding BLV users’ visual description preferences and personalized access solutions.

\subsection{Image Accessibility}
We first draw on image description literature to foreground our understanding of video descriptions. Image description guidelines (e.g., \cite{von2004labeling, mack2021designing, morash2015guiding, petrie2005describing, wcag, diagramcenter}) instruct description creators to give information about an image in relation to its context \cite{petrie2005describing, diagramcenter} and describe the predominant content (e.g., objects, people, text, scenery) to aid understanding \cite{caesar2018coco, mack2021designing, stangl2020person}. Researchers have also investigated creating descriptions through human-authored \cite{von2004labeling, burton2012crowdsourcing, simons2020hope, bigham2010vizwiz, gurari2017crowdverge} and AI-supported methods \cite{salisbury2018evaluating, salisbury2017toward, gleason2020twitter}.

To augment BLV users’ image experiences, researchers have explored methods for making images accessible beyond static textual descriptions. For example, Morris et al. \cite{morris2018rich} examined approaches for “rich” image descriptions that support interactive image accessibility, while others built systems to enable touch-based image exploration \cite{nair2023imageassist, lee2022imageexplorer}. Researchers have also studied the efficacy of using music, earcons, and tactile elements for the accessibility of artwork \cite{rector2017eyes, cavazos2018interactive, cavazos2021accessible}, museum experiences \cite{luo2023wesee, asakawa2019independent}, and data visualizations \cite{sharif2022voxlens, siu2022supporting, brown2012viztouch}. Others have researched the nuances of image consumption and visual descriptions on popular social media websites, including Twitter and Facebook \cite{gleason2019making, gleason2020making, gleason2020twitter, macleod2017understanding, morris2016most, gleason2019s, voykinska2016blind, wu2017automatic}. They found that nonverbal cues, such as short sounds to indicate the repetition of a meme format \cite{gleason2019making}, could aid in image accessibility while preserving emotion and tone.

Prior work has also emphasized that BLV people’s preferences for image descriptions vary based on an image’s context \cite{diagramcenter}. Specifically, preferences differ based on the source or content of the image \cite{bennett2021s, stangl2020person, stangl2021going, muehlbradt2022s, hanley2021computer, mott2023accessibility, chunseong2017attend, kreiss2022context, jung2022so}. For example, Stangl et al. \cite{stangl2020person, stangl2021going} found that BLV people wanted different details for images associated with different sources and user goals. Some details, such as attributes of the primary object in the image, were desired across a variety of scenarios. While some have used artificial intelligence to generate descriptions with different linguistic attributes, such as personality \cite{salisbury2018evaluating, shuster2019engaging} or writing style \cite{gan2017stylenet, chunseong2017attend}, these investigations often do not consider how a BLV user’s current context may influence their photo-viewing experience or their description presentation preferences. 

In this paper, we draw on Stangl et al.’s definition of scenarios \cite{stangl2021going} to inform our study design. We focus on contextual factors that affect BLV users’ preferences when consuming \textit{video} content and diverge from current “one-size-fits-all” models of video description to design more personalized video accessibility experiences. 

\subsection{Video Accessibility}
Audio description and video accessibility practices are guided by BLV innovators \cite{acbadpallabout}, advocacy organizations serving BLV people \cite{dcmpguidelines, descriptionkeydcmp, adcoalition, acbadpguide, dcmphowtodescribe}, and industry practitioners \cite{netflixstyleguide, 3play, fryer2016introduction}. Although a video can be conceptualized as a sequence of frames, the process of AD authorship involves additional intricacies beyond simply linking image descriptions together. Audio description requires creators to understand the video’s broader context, distill meaning from multiple frames, and insert descriptions at appropriate times within the auditory and visual narrative of the video \cite{pavel2020rescribe, zhang2023mm}. 

Some prior work in HCI has specifically explored BLV people’s audio description detail preferences. In investigations of ViScene, a collaborative AD authoring system that employed novices to increase AD availability, Natalie et al. \cite{natalie2020viscene, natalie2022cost, natalie2021efficacy} reported that BLV people valued details about clothing, time of day, and location. Additionally, through evaluations of an automated AD system, Wang et al. \cite{wang2021toward} found that BLV users preferred different details for different video types \textemdash{} for example, they wished to have more detailed descriptions of people in a comedy but not in a DIY video. Furthermore, Jiang et al. \cite{jiang2022co} found that BLV AD writers focused specifically on character descriptions (e.g., race, age), background settings, action descriptions, and clarifying audio cues, given their experience as both creators and consumers. 

Additional insights about which details, and levels of detail, that BLV users want in AD have emerged through the development of tools to streamline the AD creation process \cite{pavel2020rescribe, wang2021toward, liu2021makes, yuksel2020human, campos2020cinead, whitfield2016livedescribe, killough2023exploring}. For example, Yuksel et al. \cite{yuksel2020human} found that BLV participants had preferences for description styles and content for cooking videos, such as precise directions and accurate measurements. Pavel et al. \cite{pavel2020rescribe} developed Rescribe, an AI-supported AD tool, to investigate the viability of extended-inline AD, a new description format that looped the video’s audio beneath narration while the video’s visuals were paused. While prior works identify general guidelines for overall AD quality, we examine how BLV users’ desired details may vary for different types of videos.

Other research has shown that BLV people wish to interact and engage with video content in ways other than only listening to preset neutral descriptions during the video itself. For example, Stangl et al. \cite{stangl2023potential} and Bodi et al. \cite{bodi2021automated} investigated the viability of providing video access through interactive visual question answering, reinforcing the importance of BLV users having agency in the process of making videos accessible. Others explored the impact of changing the tone or style of verbal descriptions for select video types, finding that alternative AD styles were engaging for BLV users \cite{fels2006odd, walczak2017creative, udo2010horatio, jiang2023beyond}. Additionally, Romero-Fresco et al. \cite{romero2013could} conducted a preliminary study on the efficacy of audio introductions for providing additional detail to described films. They identified that BLV participants were in favor of accessing descriptions prior to watching a movie to improve their understanding about the characters, settings, and visual style of the film. 

Prior work has also explored using different modalities to make videos accessible. In a study with BLV AD users, sighted AD creators, and BLV AD creators, Jiang et al. \cite{jiang2023beyond} identified key video accessibility considerations posed by BLV AD creators due to their unique intersection of experiences. The authors found that the linguistic and aural presentation of AD, sound design, and multisensory aspects contributed towards greater immersion for BLV viewers. Sackl et al. \cite{sackl2020ensuring} also explored how visual enhancements such as contrast adaptation, color manipulation, and sharpness adjustments could improve video accessibility for BLV users. Others evaluated how spatial audio could augment sports broadcasts \cite{jain2023towards, jain2023front}, short films \cite{lopez2022seeing}, and 360° videos \cite{chang2022omniscribe}. We build on existing research to concretely consider how video accessibility preferences can differ across scenarios and understand how to leverage nonverbal output modalities to best suit user needs.

While most prior research on video accessibility has focused on creating universally satisfactory descriptions, some have investigated the diversity of BLV users’ preferences and information needs for the same video \cite{jiang2023beyond, chmiel2022homogenous, wang2021toward, chmiel2016researching}. For example, Chmiel and Mazur \cite{chmiel2022homogenous} examined AD preference differences between people with different vision levels and experiences with vision loss onset. They concluded that “middle-of-the-road” solutions could support the AD information needs of most users, but recommended future research to study how different levels of detail for AD could best suit individual preferences. Additionally, individual interests in spatialization and multisensory interactions can vary depending on users’ abilities (e.g., spatialized audio may not be accessible for d/Deaf and hard of hearing users) and vision levels (e.g., low vision people may not want as much AD detail) \cite{jiang2023beyond}. Despite some progress in understanding BLV users’ preferences, most work in this area does not include short-form video content \cite{nevsky2023accessibility} or consider how nuances from users’ unique video watching \textit{scenarios} may give rise to different description needs \cite{chmiel2022homogenous, natalie2021efficacy}.

Building on prior work, we explore more holistic approaches for video accessibility to address BLV users’ preferences across contexts. In this study, we focus on how a user’s \textit{scenario} impacts their video watching experiences.

\section{Formative Survey}
Before beginning our interview study, we conducted a survey to gain a preliminary understanding of BLV people’s video watching behaviors and inform our interview protocol. We also used the survey as a recruiting tool for the interview. 

\subsection{Survey Methodology}
We recruited participants through social media postings, mailing lists, and snowball sampling. Our recruitment notice indicated that participants had to be 18 years old or older, identify as blind or low vision, and have experience regularly watching online videos, which we defined as using a video platform such as a social networking service, a video sharing site, or a streaming service at least twice a week. A total of 101 respondents completed our survey, including 40 who identified as blind and 45 who identified as low vision. We discarded data from 16 people who identified as sighted. For every survey response, we donated \$2.50 to the National Federation of the Blind and \$2.50 to the American Council of the Blind. Our survey procedure was approved by the Institutional Review Board (IRB) at our university.

Our survey instrument, provided in Supplementary Materials, included both multiple choice and long answer questions and was designed to take about 10 minutes to complete. We asked respondents about which platforms they used to watch video content (e.g., Netflix, TikTok, YouTube, news sites), types of videos they watched (e.g., informational / educational, comedic, how-to, videos from friends or family), and how often they used description services (e.g., only watch videos with AD, use AD whenever available, only for certain types of videos). We also asked how video accessibility could be improved overall, what types of videos would be most useful to have AD, and if there were specific types of videos they would like to watch but are currently inaccessible.

To analyze our survey responses, we calculated the frequencies of video types viewed and platforms used by BLV participants. The first author analyzed open-ended responses and identified common video types and platforms, video accessibility ideas, AD usage patterns, and useful situations for accessible videos.

\begin{table*}
    \begin{center}
	\renewcommand{\arraystretch}{1.15}
    \caption{BLV survey respondents’ commonly viewed video types and commonly used video platforms. \# represents the number of responses and \% represents the proportion of responses (of 85 total BLV respondents). VSS = video sharing site, SS = streaming service, and SNS = social networking site.}
    \Description{Two tables side by side showing commonly viewed video types and commonly used video platforms. Columns from left to right for Table 1A: Video Type, \#, \%. Columns from left to right for Table 1B: Category, Video Platform, \#, \%.}
    \label{table:survey}
        \begin{tabular}{p{0.32\textwidth} p{0.05\textwidth} p{0.05\textwidth}}
            \toprule
            \textbf{Video Type} & \textbf{\#} & \textbf{\%} \\
            \midrule[\heavyrulewidth]
            Informational / educational & 66 & 77.6 \vspace{0.01em}\\
            Comedic & 54 & 63.5 \vspace{0.01em}\\
            How-to / do it yourself & 49 & 57.6 \vspace{0.01em}\\
            Lifestyle & 45 & 52.9 \vspace{0.01em}\\
            News / commentary & 45 & 52.9 \vspace{0.01em}\\
            Science fiction / fantasy & 44 & 51.8 \vspace{0.01em}\\
            Thriller / horror & 41 & 48.2 \vspace{0.01em}\\
            Action / adventure & 41 & 48.2 \vspace{0.01em}\\
            Music videos & 36 & 42.4 \vspace{0.01em}\\
            Videos from friends or family & 36 & 42.4 \vspace{0.01em}\\
            Romance & 34 & 40.0 \vspace{0.01em}\\
            Animation & 33 & 38.8 \vspace{0.01em}\\
            Other (sports, religious, etc.) & 26 & 30.6\\
            \bottomrule
        \end{tabular}        
    \quad \quad
        \begin{tabular}{p{0.1\textwidth} p{0.2\textwidth} p{0.05\textwidth} p{0.05\textwidth}}
            \toprule
            \textbf{Category} & \textbf{Video Platform} & \textbf{\#} & \textbf{\%} \\
            \midrule[\heavyrulewidth]
            VSS & YouTube & 71 & 83.5\\
            \midrule
            SS & Netflix & 63 & 74.1\\
            & Disney+ & 33 & 38.8\\
            & Amazon Prime & 31 & 36.5\\
            & Hulu & 30 & 35.3\\
            & HBO Max & 25 & 29.4\\
            \midrule
            SNS & Facebook & 50 & 58.8\\
            & TikTok & 42 & 49.4\\
            & Instagram & 40 & 47.1\\
            & WhatsApp & 20 & 23.5 \\
            & Reddit & 12 & 14.1 \\
            \midrule
            Other & News sites & 19 & 22.4 \\
            \bottomrule
        \end{tabular}
    \end{center}
\end{table*}

\subsection{Survey Findings and Discussion}
According to the survey, the five most popular video types were: informational / educational (77.6\% of respondents), comedic (63.5\%), how-to / DIY videos (57.6\%), lifestyle (52.9\%), and news / commentary (52.9\%). For the three categories of platforms, the most used video sharing site was YouTube (83.5\%), the most used streaming service was Netflix (74.1\%), and the most used social networking site was Facebook (58.8\%). Aggregated responses for video type and platform are presented in Table \ref{table:survey}. 

Most participants used multiple devices to watch videos. 90.6\% of respondents reported using mobile phones, 71.8\% reported using computers, and 64.7\% reported using televisions. 4.7\% used tablets, and one used a projector. 

Survey respondents most frequently mentioned the following video types as ones that acutely required better video accessibility measures: how-to, informational / educational, action, content reliant on visuals (e.g., infographics, maps, or visual effects), and news and weather. Some also wanted more accessible music videos, foreign language videos, videos with minimal dialogue or only music in the background, sports, vlogs, and live events such as theater productions and concerts.

With regards to frequency of watching videos with AD, 31.8\% of participants used AD “whenever available,” and 11.8\% of participants selected that they “only watch videos with audio description” and use description “whenever they are available.” For example, one respondent explained their reasoning: \textit{“I will sometimes watch movies or TV shows [without AD] because I have a cited [sic] partner who can assist me, but for genres like a horror [sic], we will not watch a movie if audio description is not available.”} 

However, others did not use AD as often. 14.1\% of participants reported that they only used AD for certain types of videos, in certain situations, or on certain platforms. For instance, another respondent used AD for movies, but not for \textit{“concerts because I don’t like the audio description talking in the middle of a song [or for] standup comedy because it is hard to hear the comedian talking during the audio description track.”}

Due to our relatively limited sample size, we do not draw any specific conclusions about BLV people’s video watching habits and note that further investigation is warranted. However, our survey did allow us to gain insight into the breadth of videos that BLV people watch. We utilized these survey results, specifically the types of videos that they wished could be more accessible, to guide our interview protocol. 

\section{Method}
To delve deeper into BLV people’s preferences for video accessibility across viewing scenarios, we conducted semi-structured interviews with a subset of our survey respondents. During the interviews, we asked participants to expand upon their survey responses, probed about prior experiences with watching accessible and inaccessible videos, and engaged participants in a video co-watching session to brainstorm video accessibility ideas.

\subsection{Participants}
We invited 15 BLV survey respondents to participate in our interview study. We selected participants who were at least 18 years old, identified as blind or low vision, were comfortable communicating in English, and had experience regularly watching online videos as per our survey inclusion criteria. 

All participants identified as blind (none identified as low vision), but had varying degrees of residual vision. For example, some participants had light perception and could read with magnification, others did not have light perception, and one participant was born sighted and started experiencing vision loss in his 30s. Their ages ranged from 24 to 62 (mean 39.9, SD 11.3). Nine participants identified as women, five identified as men, and one identified as agender. Some participants had extensive experience with AD: one was a hobbyist blind film critic, another worked as an AD consultant and advisor, and yet another was an extended reality sound and media artist with AD creation and production experience. All participants used screen readers and five regularly used Braille displays. 

Participant demographic details are presented in Table \ref{table:participants}. Our interview protocol was approved by our university’s IRB.

\begin{table*}[ht!]
	\renewcommand{\arraystretch}{1.15}
    \begin{center}
    \caption{We present participant pseudonyms and demographics, including gender and ethnicity in participants’ own words. All participants identified as blind; here, we paraphrase their self-disclosed vision details. We also indicate the platforms they use for video watching, their AD usage (e.g., only in certain situations, whenever they are available, only watching videos with AD), and which of the three common scenarios they watched.}
    \Description{Study participant demographics for 15 participants. Columns from left to right: Pseudonym, Age / Gender, Ethnicity, Vision Details, Platforms, AD Usage, and Video.}
    \label{table:participants}
        \begin{tabular}{>{\raggedright}p{0.1\textwidth} >{\raggedright}p{0.115\textwidth} >{\raggedright}p{0.1\textwidth} >{\raggedright}p{0.27\textwidth} >{\raggedright}p{0.11\textwidth} >{\raggedright}p{0.105\textwidth} >{\raggedright\arraybackslash}p{0.06\textwidth}}
            \toprule
            \textbf{Pseudonym} & \textbf{Age / Gender} & \textbf{Ethnicity} & \textbf{Vision Details} & \textbf{Platforms} & \textbf{AD Usage} & \textbf{Video} \\
            \midrule[\heavyrulewidth]
            Alice & 30 / Female & Chinese American & Completely blind, born legally blind and lost more vision in 20s & SS, VSS, SNS & Situationally & V1 \\
            Blair & 62 / Female & Caucasian & Born with low vision, now have light perception, limited peripheral vision & SS, VSS & Whenever available & V1 \\
            Colin & 36 / Male & White & Born with low vision and lost remaining sight at 19 & SS, VSS, SNS & Whenever available & V1 \\
            Diana & 34 / Female & Asian / Pacific Islander & Blind since birth, sees shapes and colors, can read text with significant magnification & SS, VSS, SNS & Whenever available & V1 \\
            Emily & 29 / Female & White & Has light perception, born blind & SS, VSS, SNS & Whenever available & V1 \\
            Felix & 40 / Male & Caucasian & No central vision, pockets of peripheral vision, born sighted and started losing sight at 34 & SS, VSS & Only watch with AD & V2 \\
            Grace & 29 / Female & White & Left eye nothing, can read text with significant magnification & SS, VSS, SNS & Whenever available & V2 \\
            Haley & 40 / Female & Puerto Rican & Only see color and movement, born with vision but lost gradually & SS, VSS & Whenever available & V2 \\
            Isaac & 48 / Male & White & Blurry tunnel vision, can read large print, gradual vision loss & SS, VSS, SNS & Situationally & V2 \\
            Julia & 59 / Female & White & Has light perception, no color, born with low vision, experienced significant vision loss at 28 & SS, VSS, SNS & Whenever available & V2 \\
            Karla & 24 / Female & Hispanic & Completely blind (no light perception), born blind & SS, VSS & Whenever available & V3 \\
            Layne & 41 / Agender & White & One eye with no vision, other eye has no peripheral vision or depth perception & SS, VSS, SNS & Whenever available & V3 \\
            Mason & 38 / Male & White & Completely blind (no light perception), born blind & SS & Only watch with AD & V3 \\
            Nicki & 34 / Female & Hispanic & Has light perception, cannot see shadows, lost vision over time & SS, VSS, SNS & Whenever available & V3 \\
            Oscar & 54 / Male & Caucasian & Born legally blind, gradually lost peripheral vision starting at 30 & SS, VSS, SNS & Whenever available & V3 \\
            \bottomrule
        \end{tabular}
    \end{center}
\end{table*}

\subsection{Procedure}
Our study included a virtual 75-minute interview session, conducted via Zoom. The interviews consisted of three parts: a review of the participant’s survey responses, a discussion about several recently viewed videos, and a video co-watching activity.

We first asked participants demographic questions, then reviewed their survey responses and probed about interesting comments. For example, we asked participants to expand on their rationale for only using AD in certain situations or on certain platforms.

During the second part of the interview, we asked participants to recall one or more specific videos they had watched in the last few weeks. This recent critical incident approach \cite{flanagan1954critical} allowed participants to reflect on concrete instances and provide richer, more specific insights than when speaking about general behaviors. For each video discussed, we determined the participant’s viewing scenario and asked them about positive and negative aspects of their experience. We then probed participants to explore what could make these videos more accessible, eliciting creative ideas not yet possible with current technology. Our inquiry was based on multisensory interactions for videos proposed by Jiang et al. \cite{jiang2023beyond} and Sackl et al. \cite{sackl2020ensuring}. Referring to the specific video and scenario, we asked questions such as: 

\begin{itemize}
\item How could the video overall have been made accessible, given that you are watching it in [this scenario]?
\item In what other ways would you like to have descriptions of the video?
\item Thinking about audio or sound effects more generally, what additional audio could help make the video more accessible?
\item What visual enhancements, if any, could help make the visuals more accessible?
\item What tactile feedback, if any, could help make the video more accessible?
\end{itemize}

The third part of the interview was the video co-watching portion, during which we presented participants with multiple videos to encourage them to compare and contrast their preferences \textit{across different scenarios}. To ensure that scenarios were relatable to participants and avoid a limitation of a similar study \cite{stangl2021going}, we developed naturalistic viewing scenarios based on our survey findings. The first scenario was randomly selected from three \textbf{common scenarios}, which consisted of a nature documentary, a comedy sketch, and a short-form video (more details provided in Table \ref{table:probes}). Although short-form videos were not as commonly reported in our survey, we included this video type to capture participants’ thoughts on an unfamiliar but emerging scenario. 

Next, we presented participants with one to two \textbf{participant-specific scenarios}. We pre-selected scenarios with video types that participants indicated they would like to watch in their survey responses. Our aim was to gather participants’ perspectives on how to improve the accessibility of videos that did not have any existing access measures; as such, none of the videos we presented to participants had AD. Since scenarios were assigned based on participant preferences, some video types had more responses (e.g., six participants co-watched various how-to videos while only one co-watched a foreign language film clip). However, this method allowed us to choose scenarios specific to participants’ unique viewing interests, which helped with elucidating current frustrations and brainstorming future accessibility measures. Three examples of participant-specific scenarios can be found in Table \ref{table:userprobes} and a full list is provided in Appendix \ref{appendix:allprobes}. 

\begin{table*}[ht!]
	\renewcommand{\arraystretch}{1.15}
    \begin{center}
    \caption{The three common video probes and further details, including each video’s scenario (type, platform, and user goal) and our rationale for choosing the videos.}
    \Description{Three common scenario probes used during the interview study. Columns from left to right: V1: Nature Documentary, V2: Comedy Sketch, V3: Short-Form Video. For each probe, the scenario (type, platform, and user goal) and additional details (synopsis, rationale, source, and length) are given.}
    \label{table:probes}
        \begin{tabular}{p{0.02\textwidth} | >{\raggedright}p{0.07\textwidth} | >{\raggedright}p{0.27\textwidth} | >{\raggedright}p{0.27\textwidth} | >{\raggedright\arraybackslash}p{0.27\textwidth}}
            \toprule
             & & \textbf{V1: Nature Documentary} & \textbf{V2: Comedy Sketch} & \textbf{V3: Short-Form Video} \\
            \midrule[\heavyrulewidth]
            \parbox[t]{2mm}{\multirow{3}{*}{\rotatebox[origin=c]{90}{Scenario}}} & Type & Informational / educational & Comedic & Lifestyle \\
             & Platform & Streaming service & Video sharing site & Social networking service \\
             & Goal & Learning about a concept & Entertainment & Engaging with others \\
            \midrule
            \parbox[t]{2mm}{\multirow{7}{*}{\rotatebox[origin=c]{90}{Details}}} & Synopsis & Aerial and close-up footage of wooded forests and forest floors, with narration & A man engages in a conversation with a woman at a cafe to try to guess her age & From a hotel room balcony in Paris, an influencer waves to tourists gathered outside \\
             & Rationale & Familiar video format, text on screen, sparse narration, cinematic shots & Familiar video format, text on screen, multiple characters, visual gags & Unfamiliar video format, text on screen, no dialogue, sound effects \\
             & Source & Netflix \cite{v1} & YouTube \cite{v2} & Instagram \cite{v3} \\
             & Length & 90 seconds & 90 seconds & 11 seconds \\
            \bottomrule
        \end{tabular}
    \end{center}
\end{table*}

\begin{table*}[ht!]
	\renewcommand{\arraystretch}{1.15}
    \begin{center}
    \caption{Three of our 17 unique participant-specific video probes. Videos were hosted on YouTube but represented different scenarios and platforms. The full list of participant-specific videos is provided in Appendix \ref{appendix:allprobes}.}
    \Description{Three examples of participant-specific scenario probes. Columns from left to right: Tennis Match, Cooking Tutorial, Pop Music Video. For each probe, the scenario (type, platform, and user goal) and additional details (synopsis and rationale) are given.}
    \label{table:userprobes}
        \begin{tabular}{p{0.02\textwidth} | >{\raggedright}p{0.07\textwidth} | >{\raggedright}p{0.27\textwidth} | >{\raggedright}p{0.27\textwidth} | >{\raggedright\arraybackslash}p{0.27\textwidth}}
            \toprule
             & & \textbf{Tennis Match} & \textbf{Cooking Tutorial} & \textbf{Pop Music Video} \\
            \midrule[\heavyrulewidth]
            \parbox[t]{2mm}{\multirow{3}{*}{\rotatebox[origin=c]{90}{Scenario}}} & Type & Sports & How-to & Music video \\
             & Platform & Streaming service & Video sharing site & Video sharing site \\
             & Goal & Entertainment & Learn how to do something & Entertainment \\
            \midrule
            \parbox[t]{2mm}{\multirow{3}{*}{\rotatebox[origin=c]{90}{Details}}} & Synopsis & A highlight reel of a professional women's tennis match & A video explaining how to cook four different meals & Music video for \textit{You Belong With Me} by Taylor Swift \\
             & Rationale & Minimal narration / sound & Text on screen, no narration & No dialogue, only the song \\
            \bottomrule
        \end{tabular}
    \end{center}
\end{table*}

During this part of the study, the interviewer shared their screen and audio so that participants, regardless of familiarity with the interview platform, did not have to navigate potentially inaccessible user interfaces. Prior to playing each video, we presented the participant with the scenario as follows: \textit{“Imagine you were watching a video such as [video type] on [platform]. Your goal of watching this video is [user goal].”} We played 60-90 seconds of each video, which we found in our pilot sessions to be a sufficient length of time to elicit meaningful feedback while avoiding fatigue. Participants were instructed to say “pause” if they had questions or comments about the video, but most chose not to while watching.

Each time the participant paused the video, we invited them to share their thoughts and asked about the accessibility of the video with questions such as: 

\begin{itemize}
\item What do you think just happened in the video?
	\begin{itemize}
	\item If participants had simple questions or misconceptions, we clarified by concisely providing visual information, based on guidance in prior AD work \cite{jiang2022co}.
	\item We asked participants about what they thought happened in the video before providing clarifications, as this helped identify mismatches between participants’ inferences and the video’s actual visuals.
	\end{itemize}
\item Was anything confusing or unclear? 
\item How accessible was the video so far?
\end{itemize}

Once each video was finished, we asked participants about its overall accessibility and other ways to make it more holistically accessible, using the same questions presented during the second part of the interview.

Participants were compensated with a \$30 gift card for their time and contributions.

\subsection{Data Analysis}
We audio recorded and transcribed all interviews. Three researchers analyzed the data using inductive coding. The first author individually coded two transcripts and two other authors each individually coded one of the two transcripts. Then, the authors discussed code discrepancies, developed a codebook, and split up the remaining interview transcripts for coding. After coding, three authors performed thematic analysis \cite{braun2006using} on our interview transcripts to identify overarching themes and patterns across video watching scenarios.

\subsection{Positionality}
Members of the research team identify as sighted and low vision, and have varying degrees of experience with using AD in their everyday lives. The first author, who conducted all interviews and spearheaded analysis, is a sighted person who has experience as an amateur AD creator and frequently uses AD when it is available.

\section{Findings}
Our findings are presented in terms of scenarios that participants encountered. The scenarios are anchored in video types, as we identified that video types were strongly correlated with the platform used and a user’s goals. We briefly begin each section with common barriers to watching videos in each scenario. Then, we describe the specific details, levels of detail, and output modalities that participants found helpful for accessibility. Lastly, we highlight similarities that persisted across all scenarios.

\subsection{How-To Video: Learning How to Do Something on Video Sharing Sites}
Participants frequently watched how-to videos with the goal of learning how to do something (N = 9). However, seven mentioned the lack of detailed narration as a frustrating accessibility barrier. For instance, using demonstrative pronouns (such as “this”) during narration was a primary source of confusion. Diana, who often watched knitting how-to videos, needed to search for videos that \textit{“feature[d] the person actually saying what they’re doing, as opposed to just being like, ‘And then you go like this.’”} Videos were also inaccessible when they only played music and did not include any narration at all. Karla recounted her frustrations with coming across how-to videos with background music and text on screen: \textit{“that just sounds like music to me.”} 

\subsubsection{Details about Actions and Equipment}
Ten participants explained that providing more details would help with learning how to do something. Five participants specifically mentioned that actions should be explained in \textit{“excruciatingly painful detail”} (Layne). Grace, who often watched cooking how-to videos, emphasized that such details were critical for success: \textit{“[if] this is a recipe that I would hope to replicate, I’m going to want those step-by-step very detailed instructions.”} As a former chef, Blair also enjoyed watching cooking videos. She preferred cooking videos that presented detailed information in a “technical” structure, which resembled her formal training: \textit{“[the chef] will give you your ingredients, then your method and your technique. And it's all just very logical”} (Blair). For DIY videos, Colin gave a hypothetical description which identified actions and specific corners of a piece of furniture instead of using vague instructions such as “this” or “that”: \textit{“If they are nailing nails into something, [the AD should say,] ‘He gets out three nails for this project, and nails them in corner A, corner B, and corner C.’”} Mason also wanted to know more about which ingredients and cooking utensils to use to confidently replicate a recipe. 

\subsubsection{Output Modalities}
Participants suggested various techniques to improve the accessibility of how-to videos; for example, four participants mentioned that a separate resource with additional information would be helpful. Nicki wanted to know more about products used in how-to videos, and found it helpful to include this information through narration, the video description, or comments. Similarly, when watching a home exercise tutorial, Karla thought it would be valuable to have a separate resource, such as a website or transcript with a list of the workout moves in the video.

\begin{quote}
“Having a lot more audio and textual feedback would be helpful... Having a link that you can click, or a list of different workout stuff that they're going to do, gives you some time to prep and be like, ‘Okay, now I know what I'm doing,’ so I can work out along with the video.” (Karla, 24F)
\end{quote}

Three participants also suggested that audio cues could be added to indicate a change in instruction (Karla) or a timer (Mason) to know when one step was complete and another was starting. However, Mason explained that while audio cues could be useful, their meaning needed to be properly explained: \textit{“I'd have to know what the sound is for and why. [If it’s] just a random sound, I'm just going to go, ‘That's weird,’ and ignore it.”} Emily was also interested in maintaining diegetic audio in how-to videos (e.g., the sound of an electric mixer in a baking video), as the sounds could help her determine if the task necessitated electric tools and estimate approximately how long they would need to be used.

Participants thought that having tactile information could help with their understanding as well. Mason often watched how-to videos to learn how to fix items around his home, such as a hot water heater, and wished to simultaneously read a description of the actions on a Braille display while watching the video. Additionally, Karla watched how-to videos on weaving and thought having tactile graphics throughout different stages of completion would be helpful: \textit{“having a picture of the finished product, or of the product as it’s going through [the steps] might actually be helpful.”} However, Grace cautioned against unexplained tactile cues, mentioning that a vibration would be confusing \textit{“because [she] would think [her] phone was ringing.”}

\subsection{Informational and Educational Video: Learning a New Concept on a Streaming Service or Video Sharing Site}
Informational and educational videos were present in eight participants’ viewing rotations. Alice, a student and hobbyist viewer of psychology lectures, mentioned that vague references to visual aids excluded her from forming a full understanding of educational content. Layne, who frequently watched philosophy video essays on YouTube, thought that the extensive narration common in educational videos left limited time to insert inline AD, and cautioned against descriptions that were more \textit{“disruptive”} than helpful.

\subsubsection{Details about Visual Aids, Settings, and Subjects}
Participants desired details that would help them better conceptualize abstract information. Six found that the narration and lecturing inherent to most informational videos was \textit{“accessible to a point”} (Blair); however, to support their understanding, they wished to have more information about visual aids or graphics, text on screen, settings, and subjects. Isaac described how he relied on a separate app to improve his access to infographics: \textit{“if I’m watching [a video] on my phone and it [has] an infographic, I’ll pause it, I’ll put on image recognition ... and try to see if it’ll recognize any of the text, and then try to fill in the blanks that way.”} 

During the co-watching sessions, participants’ questions during and after watching the video uncovered varied description needs. For documentaries, the setting was important to participants. When asked about what happened in the video probe, Colin mentioned how the narration helped him \textit{“get the context that it’s large trees in the dark environment”} but wished to know \textit{“the finer details, like how large the trees might be.”} Blair was also curious about \textit{“what kind of trees”} were being shown, and did not want details about clothing unless they were relevant to the educational aim of the video. Emily highlighted the value of specific descriptions for learning more about species in a documentary with the following example: 

\begin{quote}
“Instead of just saying ‘a blue bird,’ maybe say its size and beak size and wingspan... If the documentary [showed] the differences between male and female birds, a sighted person who watched that five minutes ago could now tell that a male bird and a female bird are flying toward each other. But if the describer just said ‘two birds flying towards each other,’ that’s not going to work.” (Emily, 29F)
\end{quote}

\subsubsection{Output Modalities}
As with other scenarios, six participants were open to adding output modalities to supplement their video watching experience.

Participants thought audio cues would help with video comprehension. They appreciated that ambient sounds in the video’s original audio could convey the mood and context of the documentary, which gave them \textit{“more of a feel for the environment”} (Colin). Colin was also open to learning more through \textit{“an audio track or an alternate link to click to for more information.”} Three participants proposed additions to the soundscape, suggesting that scene changes could be cued with a sound effect that was distinct from other noises in the scene, so as not to \textit{“blend in with the birds of the movie”} (Emily). 

Three participants specifically mentioned tactile graphics as a helpful tool for understanding concepts such as scale or structure. For example, Colin wished to have \textit{“[embossed] pictures of those forests”} from the documentary, while Grace thought \textit{“having tactile versions of the grid or the map would be super cool.”} However, others found tactile graphics to be hard to interpret, and instead preferred 3D models to indicate what \textit{“the animals, or... some of the rocks, or what the soil would feel like”} (Emily). While participants differed on which tactile modalities they wanted, they agreed that tactile elements were broadly helpful for learning. 

\begin{quote}
“It would be cool to have it be more tactile because that's how you learn. Descriptions might be great, but describing a part of a cell is not as good as seeing a picture or feeling it more hands-on.” (Alice, 30F)
\end{quote}

\subsection{Short-Form Video: Engaging with Friends and Pop Culture on a Social Networking Site}
Not all participants were active social media users, almost always due to the inaccessibility of platforms like Instagram and TikTok. All participants were aware of the short-form video format, and five were frequent viewers. Seven participants noted that the prevalence of text on screen (e.g., a product review video with product details listed as text on screen) was a major barrier to understanding and engaging with Instagram Reels, TikToks, and Instagram Stories.

\subsubsection{Details about Subjects, Actions, Clothing, and Settings}
Five participants were primarily interested in knowing more details pertaining to subjects (both people and pets) and their actions. Common questions that arose when discussing short-form videos included “characters” and their actions. For example, when watching an animal video, participants expressed that they at least wanted to know, \textit{“What is the animal? And what are they doing?”} (Layne). 

While this information about “who” and “what” was critical, three participants wanted details about clothing and settings. For example, Karla felt that she required \textit{“a full picture of the surroundings and the clothing”} to completely understand the context, given the short video duration. As someone who was unfamiliar with this video format, Mason also emphasized the importance of these details for understanding the short-form common scenario (V3).

\subsubsection{Output Modalities}
Seven participants wished to access additional detail or context in the video caption or through an external resource. Though AD is typically synchronous with a video, some wished to know more \textit{before} watching. Five participants often used the video caption (a text description posted by the video’s original creator) as a tool that could be referenced to \textit{“put [the video] into context”} (Layne), helping viewers infer the overall tone of the piece. Karla explained how even short captions helped her form an idea of what a video may involve:

\begin{quote}
“[If] the caption is ‘girls’ night,’ you can kind of guess what’s happening \textemdash{} you’re probably going to go out shopping or stay in and watch a movie... If the caption is ‘When girls’ night goes terribly wrong,’ you can assume that they were relaxing and then something happened.” (Karla, 24F)
\end{quote}

However, others such as Grace reported a lack of connection between the caption and the actual content of the video. She mentioned that \textit{“[creators] don’t actually say what happens in the video... the caption might be like, ‘Oh, I was so shocked’ or a bunch of hashtags”} (Grace). In these cases, she was frustrated about taking extra time to read a caption that did not provide the context she wanted. 

Audio, such as background music, could also improve accessibility. Karla and Nicki mentioned that the practice of using popular “sounds” as templates for short-form videos helped their understanding. As with captions, they found that \textit{“certain music... could tell a little bit about the video”} (Nicki), including the video’s tone and content. Karla explained how familiar music helped her infer the tone of TikToks:

\begin{quote}
“If people use [music] in the right context, some people will put the ‘oh no’ song... because there’s something that makes you go, ‘Oh no.’ ... Sometimes if it’s more of a slower piano beat, I’m like, ‘Okay, it’s probably something sad, or it’s something serious.’ But if it’s really fast, kind of a bouncy type thing, I’m like, ‘Okay, it’s probably something intended to be a little bit lighter.’” (Karla, 24F)
\end{quote}

For short-form videos, three participants generally thought descriptions were more valuable than additional output modalities for providing context. However, some thought having \textit{“different audio cues for different environments”} (Karla) could reduce the amount of information conveyed verbally. Oscar also shared how audio cues could help with his frustration about not knowing when a short-form video was automatically replaying. Tactile cues were also useful; for dancing videos on TikTok, Karla suggested having \textit{“a pattern on the Braille display that moves along with the dancing [movements]”} to better convey the action in a nonvisual way.

\subsection{Music Video: Seeking Entertainment on a Video Sharing Site}
\begin{quote}
“I’ve heard this song many times... but I’ve never really known what happens exactly in the video. How does it start? ... Who’s in the video? ... What’s happening in the story?” (Nicki, 34F)
\end{quote}

Watching music videos for entertainment was a unique scenario, as participants were often familiar with the music itself but were excluded from enjoying the stories in the accompanying videos. In fact, two participants shared that they were uncertain if music videos contained visuals at all: \textit{“all I’m hearing, as a blind person who can’t really tell what’s happening on the screen, is the music. It’s almost like I’m just playing the song without any context”} (Nicki). Four other participants attempted to infer action based on the video’s audio or comments, often with limited success.

\subsubsection{Details about People, Actions, Settings, Clothing, and Visual Effects}
Especially for familiar songs, participants were interested in having access to details that gave them an idea of the story presented through the video. Five participants primarily wished to know who was present in the music video and what they were doing. For example, Nicki mentioned how the song lyrics in Taylor Swift’s song, \textit{You Belong With Me}, suggested that the video could contain certain characters (e.g., a cheerleader). However, she was curious about which other characters were in the scene and wished to know more about what the characters were wearing, as \textit{“sometimes what people are wearing can tell you a little bit more about the context”} (Nicki). Similarly, Blair wanted to have enough information to \textit{“use [her] imagination and make pictures for [herself].”} 

\begin{quote}
“I want to know what the outfits are. I want to know the dancing or the setting, the scenery. I want them to set the stage for me \textemdash{} really, literally set the stage for me \textemdash{} the hair, the makeup, everything.” (Blair, 62F)
\end{quote}

Lastly, Alice felt \textit{“privy”} to details, including visual effects or flashbacks, that conveyed the cultural and political commentary behind a music video. For example, she recalled wanting to know more about the political context behind a controversial country music video. However, because the video lacked descriptions, she found it difficult to participate in broader discourse about the video. 

\subsubsection{Output Modalities}
Participants shared a myriad of ways to make music videos accessible beyond traditional AD methods. Two participants preferred inline descriptions, and two others wished to have extended descriptions as their goal of understanding the story \textit{“scene by scene [without feeling] pressed for time”} (Nicki) was prioritized over listening to the music itself. One of the inline AD and one of the extended AD advocates recommended having a separate resource to reference for more details, such as a descriptive “prologue” to set the scene of the music video prior to watching it. 

Regarding how the descriptions interacted with the music, two participants viewed music and lyrics to be different from dialogue. They suggested that the music could be ducked or even omitted in favor of AD during repetitive sections such as the chorus.

Three participants were also interested in tactile elements such as Braille and haptics, preferring tactile cues over additional audio. Emily explained that Braille could be helpful for providing AD during musical sequences, noting that audio ducking requires producers to \textit{“turn the song down”} whereas Braille could allow users to hear the song while getting descriptions. Additionally, Isaac thought haptic feedback could improve his immersion with a music video. For example, he suggested: \textit{“when somebody gets punched, your phone can vibrate... just to amplify the experience, especially if you’re watching it on your phone where you don’t have the subwoofer.”} 

\subsection{Live Video: Seeking Information or Entertainment on a Video Sharing Site or Social Networking Site}
Eight participants reported that they watched live videos, such as news, sports, and live streams, to seek information or be entertained. Though these videos typically involved live narration, making them somewhat accessible, they often lacked detailed visual descriptions compared to videos with AD added in post-production.

\subsubsection{Details about Visual Aids, People, Actions, and Clothing}
For news, descriptive details about visuals were crucial for participant safety. Such visual information applied to infographics as well as live-action clips. For example, during a weather broadcast, Grace was frustrated by usage of generic demonstrative pronouns (such as “this”) to refer to specific locations on the weather map, as impending weather conditions could require viewers to take action.

\begin{quote}
“I struggle with maps [when] a newscaster [says], ‘it [will] rain and it’s going to go this way.’ I don't know if that's hitting near me. So I like it when they talk through a map. But in general, it can be hard because they’re just making blanket statements about a state. And you don't know like, ‘Okay, is it coming closer to where I am?’” (Grace, 29F)
\end{quote}

Blair was particularly interested in live sports. While she sometimes listened to radio coverage of sports to get more description, for videos she desired \textit{“play-by-play”} (Blair) commentary and wished to know more details such as the players’ clothing, actions, and facial expressions. She explained that these details were \textit{“all part of the anticipation of tennis”} (Blair). 

\subsubsection{Output Modalities}
Audio cues could also indicate a scene change during news broadcasts. For example, Emily wished to have AD for news footage, such as overhead shots of scenery, but noted that an audio cue would suffice for indicating the program had switched back to a shot of the news anchor in the studio. 

Unlike non-interactable news and sports broadcasts, online platforms such as Facebook, YouTube, and Twitch support live streams where creators can directly respond to BLV viewers’ requests. For example, Haley recounted her experiences of asking questions during a kitten sanctuary live stream. While the live streamers often described what the kittens were doing, Haley sometimes utilized the chat to request that they \textit{“describe the kittens themselves.”} She found that the creators and other viewers in the chat usually responded positively to her request for access in real time. 

\subsection{Personal Video: Engaging with Friends or Family on a Social Networking Site}
Many participants used social media, most often Facebook, to connect with and watch videos from friends and family (N = 6) or for entertainment (N = 4). As such, video subjects often included people or pets that the viewer knew personally, and the content ranged from pets playing to children’s recitals. Participants reported that these videos were most often inaccessible when they did not include much speech or dialogue.

\subsubsection{Details about People, Pets, Actions, Settings, and Clothing}
Participants wished to know more about people or pets and action. Three participants emphasized that the people or pets in the video were most important to describe: \textit{“if someone is showing a video of their cat, they first want you to focus on their cat and what their cat is doing”} (Emily). Similarly, Nicki emphasized that knowing about the people helped her stay updated with her family: \textit{“for example, if there’s a baby in the family, [I want] to see their progress, how much they’re growing, what they are accomplishing.”} 

Three participants were also interested in settings and clothing, but noted that this was less critical. For videos that already implied the people and actions through dialogue, describing the setting was helpful. For example, Felix recalled watching a video of his friend’s child performing in a play, and mentioned that he wanted AD to give \textit{“the context of, where’s this person singing? What’s on stage? What’s the setting here?”} Nicki explained that finer details about clothing and colors helped her feel more connected to the emotions captured in the videos. 

\begin{quote}
“Maybe someone just got their hair done or dyed, and I want to know what color they dyed their hair... If it’s a video of someone at the beach, I want to know what color the water looks like, I want to know, is the sky sunny or cloudy? Things like that still bring joy to me.” (Nicki, 34F)
\end{quote}

\subsubsection{Output Modalities}
For videos from friends and family, participants did not expect nor want professionally produced descriptions or output modalities. Instead, they wished for the videos to have descriptive narration during the actual filming process of the video, such as \textit{“saying, ‘Oh, we’re at our local park and so and so is going down the slide for the first time’”} (Nicki), or through accompanying context clues in the video caption. Karla also mentioned that her friends and family would sometimes preface a video’s content when sharing it as a proactive way to provide access. For times when friends or family forgot to provide additional context, Emily suggested that platforms could \textit{“remind people to add description.”} 

\subsection{TV Show or Movie: Seeking Entertainment on a Streaming Service}
Across multiple genres of television and movies, a majority of participants (N = 11) were interested in knowing more about characters’ appearances, actions, clothing, facial expressions, and settings, findings that are in accordance with most existing AD guidelines. However, participants’ preferences also illuminated differences across genres, most commonly through their proposed output modalities. Here, we present participants’ suggestions in groups based on similar findings.

\subsubsection{Science Fiction, Fantasy, and Animation}
\begin{quote}
“There’s something about \textit{Wall-E}, which is one of my favorite films of all time, that just does not translate... As somebody who could see \textit{Wall-E} and now cannot, I can tell you that the audio description just doesn’t have it. It tries really hard to capture the magic, but there’s something that is missing out of the little expressions that the characters have that’s so hard to describe.” (Felix, 40M)
\end{quote}

Genres with fantastical elements, such as science fiction, fantasy, and animated content, were often difficult to describe within the constraints of dialogue gaps. Felix, a blind movie critic who lost the majority of his sight at 34 years old, was especially emphatic about how separate resources and additional output modalities could help with his understanding and enjoyment. Separate resources were valuable for including detailed explanations of characters and clothing, such as \textit{“what a stormtrooper wears”} (Felix). He highlighted the impact of having a prologue to set the scene of a show:

\begin{quote}
“I’ve seen films that are so immers[ed] into a fantasy or sci-fi realm... where nothing has a basis in reality... We could have an additional... immersive audio description [prologue] to describe the world in which we’re about to live. ... That way we can really focus on the story, the characters, the plot, the relationships \textemdash{} what’s happening.” (Felix, 40M)
\end{quote}

Having physical 3D models could also assist in conveying the unique designs and nonverbal expressiveness of animated characters. When discussing animated character design, Felix mentioned: \textit{“I know we can’t roll a Wall-E into people’s homes, but I almost wish we could.”} Emily was also a staunch supporter of utilizing 3D models for video accessibility, mentioning that having access to a mermaid doll helped her conceptualize what Ariel from \textit{The Little Mermaid} looked like. Regarding the practicality of actually obtaining these 3D models, she acknowledged that not all viewers would have access to 3D printers. To make this more feasible and to reduce the technical overhead, she suggested implementing \textit{“a rental program in libraries [such as] the National Library Service for the Blind”} (Emily) or making the 3D models available at movie theaters. 

\subsubsection{Comedy}
Five participants prioritized details that could provide access to \textit{“sight gags”} (Isaac), such as humorous text on screen, facial expressions, or even graphic clothing that contributed to punch lines. Grace recalled her experience watching the show \textit{Lizzie McGuire} and how \textit{“a little cartoon would come in a thought bubble and say her thought[s],”} suggesting that a similar \textit{“cartoon guy voice... [with] rising inflection on the end”} could not only read out text on screen, but also present it in a humorous way. 

Others stressed the importance of ensuring that sound effects, both in the original work and added for accessibility purposes, conveyed the right tone. When referring to the video probe from the interview (V2), Isaac mentioned how the video successfully indicated that the ticker on screen was going up and down: \textit{“the slide whistle was very, very helpful... the Foley in the show itself did a good job of describing what was going on.”}

Three participants thought tactile elements were not necessary for comedy videos. However, they did mention that it was especially important to not over-explain a joke or prematurely spoil the humor. When talking about the timing, Julia advocated, \textit{“I’d want to go along with it as it’s happening in order to make it really funny and really entertaining... I want to be in the moment with everything else.”}

\subsubsection{Historical, Romance, Reality, and Drama}
Six participants watched historical, romance, reality, and drama content for entertainment and relaxation. For reality shows, two participants shared that they were often interested in clothing for the sake of getting a better understanding of a person’s character. In particular, when watching a reality dating show, Diana mentioned how her wife would often pause the show and describe a \textit{“ridiculous bathing suit”} or other visual details ad hoc, in a more subjective manner than traditional AD. While she did not want to spend too much time familiarizing herself with a show she had just started watching, she was interested in having a separate resource to reference once she became invested with a show. She recounted her positive experiences with sharing these additional descriptions with friends:

\begin{quote}
“I actually typed up [my wife’s] descriptions of everybody, and sent it to my other blind friends who are watching the show. They were like, ‘Oh my god, this is so great.’ I wish there was a thing where, on shows like this, you could choose to go in and access this additional description... because a lot of the time it’s just not feasible to put that in there. And obviously, that is kind of subjective on some level, so maybe that’s not something that a production company would feel comfortable providing, but it really does enhance the experience to know: is this person who acts really vain actually super hot?” (Diana, 34F)
\end{quote}

Regarding character appearances, Felix mentioned the harms of omitting race and ethnicity from AD: \textit{“a person’s race is not revealed unless it’s not white, and so we’re all left to assume that unless told otherwise, everybody is white.”} While not all participants were interested in clothing and appearance, they generally appreciated if AD could include race, citing shows such as \textit{Bridgerton} as a good example of how describing characters’ races could highlight equitable representation in media. 

\subsection{Similarities Across Scenarios}
While participants generally preferred different details for different scenarios, with Layne even commenting that for \textit{“a lot of internet media, it’s kind of context dependent,”} some recurrent suggestions illuminated universal video accessibility needs. Many of these similarities echoed existing guidelines, but participants also contextualized how these ideas could be helpful for emerging scenarios. For example, having access to text on screen was critical for accessibility. Most thought providing text-to-speech or screen reader functionality for text on screen could be helpful, but others cited their disdain for \textit{“the TikTok automated voice”} (Layne) and wanted a human narrator instead. Six participants also wanted diegetic audio to contextualize a video and alleviate the need for verbal descriptions, and ten suggested adding new audio effects for increased comprehension. However, some participants expressed concerns about augmentative audio overshadowing the creative vision of the original video: \textit{“at some point I wonder, is that people messing with it, or what the creator had in mind?”} (Alice). Four participants also wanted transcripts, and two others specifically advocated for dubbing for foreign language videos. 

Others suggested adapting a video’s visual style to enhance accessibility. Five participants with residual vision found competing colors, flashes, and fast action detrimental when watching videos, and wished to have increased contrast, \textit{“minimal background[s]”} (Grace), and less rapid action to make videos easier to comprehend. Isaac, who found \textit{“flat and simple”} animation styles in shows such as \textit{The Simpsons} easier to view, shared his enthusiasm for changing the stylistic appearance or \textit{“stripping some of the detail”} of a video: \textit{“if you’re able to choose a filter and what these characters could look like based on your vision and the way you would prefer to see things... that would be cool.”} 

Though our results focus more on the diversity of BLV people’s perspectives, rather than the generalizability of insights across scenarios, we present some similarities regarding desired details and output modalities in tabular form in Appendix \ref{appendix:summary}.

\section{Discussion}
Our findings detail how different scenarios give rise to varied video accessibility needs. BLV users were in favor of both verbal descriptions and nonverbal output modalities, such as audio cues to indicate scene changes for news, tactile elements to give a sense of character design for science fiction movies, or visual enhancements to increase contrast for fast-paced videos. Though most prior research on image and video descriptions has focused greatly on one outcome for end users, we build on work by Stangl et al. \cite{stangl2021going} to go beyond universal design and “one-size-fits-all” descriptions. To our knowledge, we are the first to explore varied preferences for video accessibility across a wide set of scenarios. 

As technology advances, viewing habits change, and content evolves, it is essential to break away from only adding AD based on traditional guidelines and consider more holistic video accessibility. During our study, participants mentioned a variety of enhancements, ranging from 3D models for unfamiliar concepts to additional resources providing detailed descriptions. The myriad of ideas shared by participants illuminates an emerging design space with more depth and breadth than current AD practice. 

In this section, we define a video accessibility design space to provide video creators and video platform designers with an expanded toolkit for making videos more holistically accessible. We then discuss the potential and implications of generative AI’s applications to video accessibility and personalization.

\subsection{Defining the Design Space for Video Accessibility}
Prior HCI work on video accessibility has focused predominantly on providing universal access through audio description \textemdash{} concise, objective narrations spoken during gaps in dialogue. However, since the introduction of AD between the 1960s and 1980s, the video watching landscape has shifted dramatically. Now that we often view videos on our personal devices, many of which are handheld, videos and devices can simultaneously provide other types of output, including haptic and tactile feedback. Additionally, as shown through our findings, participants were interested in using augmentative outputs to convey information for different video scenarios. Our notions of video accessibility should expand to consider the affordances of today’s video viewing scenarios and the full context of a user’s experience.

Below, we distill the ideas mentioned by participants into six continuous or categorical dimensions to articulate a design space for video accessibility. We also present the design space in Table \ref{table:designspace}. The first two dimensions are grounded in traditional AD practice \cite{acbadpguide, dcmpguidelines, adcoalition, descriptionkeydcmp}, while the other four are not yet as common. Similar to other design spaces, these dimensions represent an infinite possibility of different video accessibility solutions. 

\begin{table*}[ht!]
	\renewcommand{\arraystretch}{1.15}
    \begin{center}
    \caption{Our six-dimensional video accessibility design space.}
    \Description{Six dimensions, where the first three are continuous and the last three are categorical. Endpoints or examples are listed for each dimension. The dimensions are as follows: Level of Detail, Alteration of Video Time, Level of Augmentation, Modality of Presentation, Synchronicity of Accessible Content, and Tone and Style of Approach.}
    \label{table:designspace}
        \begin{tabular}{>{\raggedright}p{0.02\textwidth} | >{\raggedright}p{0.28\textwidth} | >{\raggedright\arraybackslash}p{0.64\textwidth}}
            \toprule
              & \textbf{Dimension} & \textbf{Endpoints / Examples} \\
            \midrule[\heavyrulewidth]
            \parbox[t]{2mm}{\multirow{4}{*}{\rotatebox[origin=c]{90}{
            Continuous}}} & Level of Detail & Minimal detail (concise) $\xleftrightarrow{}$ extreme detail (verbose) \\
             & Alteration of Video Time & No increase to source material duration $\xleftrightarrow{}$ extension of source material duration \\
             & Level of Augmentation & No accessibility measures added $\xleftrightarrow{}$ any number of accessibility measures added after a video’s initial production or release \\
            \midrule
            \parbox[t]{2mm}{\multirow{4}{*}{\rotatebox[origin=c]{90}{Categorical}}} & Modality of Presentation & Spoken descriptions, audio cues, visual enhancements, Braille, tactile graphics, 3D models, haptics, etc. \\
             & Synchronicity of Accessible Content & Before, during, or after watching a video \\
             & Tone and Style of Approach & Excited, sad, judgmental, first-person perspective, cartoonish, etc. \\
            \bottomrule
        \end{tabular}
    \end{center}
\end{table*}

\subsubsection{\textbf{Level of Detail (continuous)}: minimal detail (concise) $\xleftrightarrow{}$ extreme detail (verbose)}
To create more effective baseline descriptions, video accessibility creators can leverage known intrinsic qualities of the video, such as the video’s type and generally the video’s platform, to determine which details to include. Though user goals are often tied to the type and platform, different users may watch the same video for a variety of purposes, and detail preferences vary between BLV users. Our findings can serve as a guide for AD creators to determine which details are of particular interest to BLV audiences for different video types and platforms. 

However, given that user goals vary across scenarios, we emphasize that there is no singular “ground-truth” \textemdash{} detail levels should be dynamic and personalizable. BLV users should be able to indicate if they want more or less AD detail through a slider or menu. This setting could be saved universally for a user, but more ideally, should be variable across scenarios (e.g., a system should consistently provide many details for a how-to video but few details for a science fiction film if those are the detail levels a user indicates). Prior work has emphasized the importance of personalizable settings for closed captioning to direct focus and avoid distractions \cite{gorman2021adaptive, cavender2009classinfocus, jain2018towards}; we recommend for video accessibility systems to also support high degrees of flexibility. 

\subsubsection{\textbf{Alteration of Video Time (continuous)}: no increase to source material duration $\xleftrightarrow{}$ extension of source material duration}
While inline descriptions, which require no increase to source material duration, are the standard in the AD industry, some audio describers have chosen to extend the duration of source material for particular video types (e.g., fast-paced movie trailers \cite{spidermanextended}) to provide more time for AD delivery. For short-form videos such as TikToks or Instagram Reels, participants generally wanted additional descriptions to augment the limited AD given during a video’s restrictive dialogue gaps. However, while extended AD may allow for the inclusion of extra information, video creators should also consider the additional time and labor that BLV people may have to incur as a result \textemdash{} even a 30-second extension may double the amount of time a BLV person spends on a video. 

\subsubsection{\textbf{Level of Augmentation (continuous)}: no accessibility measures added $\xleftrightarrow{}$ any number of accessibility measures added after a video’s initial production or release}
Most videos were found to benefit from some degree of augmented accessibility measures. Some videos are already accessible by nature, meaning that the source video’s audio inherently conveys some information to the user through dialogue, narration (including voice-over narrations added during the editing process), diegetic audio, or other audio effects. Other videos are made accessible afterwards through the addition of AD. Participants also referenced videos that were completely inaccessible (e.g., how-to videos with only music in the background), which required multiple augmentations for complete access. The degree to which a video is understandable from its audio can help determine how extensively to augment the video, which can be through AD and / or other modalities.

\subsubsection{\textbf{Modality of Presentation (categorical)}: spoken descriptions, audio cues, visual enhancements, Braille, tactile graphics, 3D models, haptics, etc.}
Additional audio elements can improve video accessibility. Similar to how Netflix’s title card includes a recognizable sound effect \cite{tudum}, different platforms may adopt a set of distinctive earcons to efficiently indicate common information. For example, streaming services may wish to standardize a sound indicating that credits are rolling, while content creators may wish to designate a specific earcon for encouraging viewers to “subscribe” or “follow” their content. Video creators can reference podcasts and radio sportscasts as strong examples of descriptive and rich audio experiences that leverage vocal performance and sound design to engage listeners \cite{kleege2023fiction}, and may also draw on prior research exploring how sound design can enhance the accessibility and aesthetics of auditory websites \cite{zhang2022exploring}. Some films have already adopted such practices of using sound design techniques (e.g., sound effects, 3D audio) to enhance the experience for BLV audiences \cite{yorkbafta, picturelessfeature}. 

Prior research has found that augmentative tactile elements are helpful for improving access to artwork \cite{cavazos2018interactive, cavazos2021accessible, stangl2019defining} and theater experiences \cite{udo2010enhancing}. We found that participants wanted tactile feedback in formats that were not available through just a smartphone \textemdash{} they also wished to have additional materials such as 3D models and Braille output. However, many participants acknowledged that this relied heavily on having (1) the requisite technology, such as 3D printers or refreshable Braille displays, (2) Braille literacy, and (3) tactile graphicacy. 

However, it is important to consider that not all participants were interested in additional output modalities \textemdash{} some wished to use them part of the time, some thought they would be easily confused with other cues such as their phone ringing, and some acknowledged the high learning curve. We recommend for audio cues to be included in a tertiary audio track to allow BLV users full control over whether and when they would like to hear audio cues in addition to AD. For emerging devices, such as extended reality (XR) headsets, video accessibility creators can borrow from prior literature in XR accessibility and game design (e.g., \cite{jiang2023beyond, kruijff2015enhancing, cardoso2020sensory}) to determine if additional modalities such as smell or taste are appropriate for information presentation.

\subsubsection{\textbf{Synchronicity of Accessible Content (categorical)}: before, during, or after watching a video}
Primers or prologues were particularly helpful to access prior to watching exercise videos, music videos, and content with fantastical characters or extensive world-building (e.g., science fiction, fantasy, historical fiction). However, for unexpected parts of a video, such as a surprise character appearance, it was more favorable to access additional descriptions afterwards to avoid spoiling the surprise. Additionally, some participants only wished to reference separate resources and descriptions after becoming invested in a show. This extends findings from preliminary research on the positive impact of having audio introductions for select feature films \cite{romero2013could} and prior work on image exploration, which found that BLV users frequently accessed overview menus at the beginning and end of exploring images \cite{nair2023imageassist}.

\subsubsection{\textbf{Tone and Style of Approach (categorical)}: excited, sad, judgmental, first-person perspective, cartoonish, etc.}
While most existing AD is generally presented in a neutral, third-person tone \cite{dcmphowtodescribe, machuca2022neutral}, researchers have found that changing the tone or style of verbal description presentation can improve viewers’ immersion in a video \cite{fels2006odd, jiang2023beyond, walczak2017creative, udo2010horatio}. We explored how this could extend to a variety of different scenarios. For example, one participant wished to have subjective and somewhat judgmental descriptions for reality television, while others wished to have descriptions that matched the tone of the piece overall. Aside from spoken descriptions, other output modalities could also take different tones and styles; for example, tactile graphics and 3D models could present users with a more cartoonish representation of a video’s visuals to abstract away unnecessary details. 

\subsection{Design Recommendations and Examples}
Across the design space, our study yielded several recommendations for current scenarios.

\begin{itemize}
\item Video creators should provide outside resources for BLV audience members to refer to. This can take the form of a vlogger posting a visual description of themself or a blockbuster movie providing a detailed introduction to their characters. 
\item Entertainment videos (e.g., music videos, historical shows, reality television, etc.) should focus more on describing subjects, such as people and animals, which can be done through a variety of modalities. Additional information, including costumes, can help contextualize scenes, especially those with any fantastical or historical elements.
\item On the other hand, for a how-to video, subject appearances are less important. Precise information about actions and equipment should be prioritized to aid in the goal of learning how to do something.
\end{itemize}

Different points in this design space have been lightly explored in real-world settings. For example, one short-form video creator went viral for creating AD in a narrative and poetic style. When a TikTok user requested for other social media users to provide access to the videos and memes regarding the Montgomery Riverfront brawl in August of 2023 \cite{tiktokmontgomery, montgomerybrawl}, one of the responses he received was particularly distinctive for its creative and calming description of the chaotic event unfolding from the perspective of the co-captain’s hat \cite{tiktokmontgomeryresponse}. In an interview by the Washington Post, the responder, a sighted content creator, noted that \textit{“Slater’s request inspired him to try a kind of oral storytelling that transcended sensory experiences, in the style of a folk tale”} \cite{montgomerywapo}. Additionally, a Netflix original series, \textit{All the Light We Cannot See}, was one of the first television shows to be released with an official “audio introduction” \cite{allthelightintro}. The show, starring a blind lead actress and telling the story of a blind French girl and a German soldier during World War II, leveraged the audio introduction to describe character appearances, clothing, movement styles, and settings. These examples demonstrate the success of two newer dimensions, reinforcing the viability of this design space for different naturalistic scenarios.

As video viewing scenarios continue to change, new preferences may arise. We encourage future work to innovate new techniques for description presentation, explore various detail levels, and consider novel user interface designs that enable the personalization of video accessibility.

\subsection{Applying Generative AI to Explore the Video Accessibility Design Space}
To move towards user-centered and holistic video accessibility, we propose leveraging generative AI to explore the design space for different scenarios. Generative AI can serve as a tool for designers, creators, and end users to adjust video accessibility on-demand. 

Some research has focused on the development of datasets and NLP techniques for video understanding and accessibility \cite{tariq2015feature, han2023autoad, han2023autoadii, yao2015describing, youtube8m, zhou2018towards}. Others have developed AI-based tools to support accessibility practices \cite{pavel2020rescribe, wang2021toward, liu2021makes, yuksel2020human, campos2020cinead, bodi2021automated, stangl2023potential, zhang2023mm}. Major advancements in multi-modal language models such as OpenAI’s GPT-4V \cite{gpt4, gpt4v} and Google’s Gemini \cite{gemini, geminireport} show that AI is already capable of generating image descriptions, and some video descriptions, that attain high levels of quality \cite{zhang2023mm} and BLV user satisfaction \cite{bemyai2, stangl2023potential}. However, prior efforts do not specifically consider short-form video, and they primarily aim to automatically generate text descriptions (e.g., \cite{fernandez2015text, campos2020cinead}). 

To effectively train AI models for these varied use cases, it is crucial to create comprehensive datasets that reflect a wide range of scenarios, information display preferences, and output modalities. If these datasets will contain sensitive information pertaining to BLV people \textemdash{} for example, to capture the scenario of watching videos from friends and family \textemdash{} we must also recognize and consider BLV users’ visual privacy concerns \cite{stangl2022privacy, gurari2019vizwiz, stangl2023dump, zhang2023imageally}. 

Our study evidences the importance of scenario-based approaches \cite{carroll2003scenario} for video accessibility; however, given the size of the design space, it is infeasible to thoroughly explore all possible designs. As such, below we address current capabilities and limitations of AI systems for three dimensions and suggest improvements for the future of video accessibility. 

\begin{itemize}
\item \textbf{Level of Detail}: Prior work has investigated the potential for visual question answering systems to enable users to query for details that they wish to know \cite{bodi2021automated, stangl2023potential, jiang2022co}. As AI advances, it may one day be possible to provide end users with high degrees of flexibility for which details and what level of detail they would like through automatically generated descriptions. While participants did not want completely AI-generated descriptions for professionally produced content, they thought it was desirable for user-generated content created with little to no budget, such as TikToks, Instagram Reels, and videos shared by friends and family.

\item \textbf{Modality of Presentation}: Our findings show that having additional modalities for conveying visual information, such as 3D models or Braille, were welcomed by participants. Existing open-source resources and workshops are an excellent starting point for becoming familiar with tactile graphics and objects \cite{marconiussvg, dimensionschancey}. However, given the large amount of possibilities for modality, which include tactile graphics, 3D models, and audio cues, we suggest that generative AI can be valuable for quickly prototyping a wide variety of modalities to determine which would be the best fit for a specific scenario. For example, some participants wished to know the scale of the forest when watching the nature documentary. Some suggested having a tactile graphic, whereas others found 3D models more helpful for learning. Building on prior tactile graphics and tactile display research (e.g., \cite{panotopoulou2020tactile, holloway2022animations, mukhiddinov2021systematic}), non-technically savvy users could utilize generative AI to create scalable vector graphic files for tactile graphics and produce STL code for 3D printing.

\item \textbf{Tone and Style of Approach}: One of the strengths of current AI systems is their ability to mimic existing writing styles and tones \cite{gan2017stylenet, chunseong2017attend}, a capability that has been applied to both fiction (e.g., \cite{vergeai}) and scientific writing (e.g., \cite{hill2023chat}). Aside from changing the style of textual descriptions, tones and styles can also be altered for other output modalities. As some participants with residual vision mentioned, fast-moving visuals were often inaccessible, and some wished to change the visual style of the entire video. To cater to individual users’ levels of vision or stylistic preferences, generative AI can aid video accessibility creators and consumers in iterating upon different options to find the tone or style that works best for them.
\end{itemize}

Recent advances in AI models have led to excitement about potential uses of AI for video accessibility. However, we also recognize that there can be significant ethical harms associated with AI-generated and personalized video accessibility, and we caution against the unregulated deployment of such technologies. Currently, BLV people typically receive AD from a trusted friend or a company that undergoes multiple iterations of quality control. If video accessibility becomes largely automatically generated, and if there are limited methods for assessing the quality of the output, the impacts of biases and misinformation perpetuated by AI can become magnified \cite{whittaker2019disability, ji2023survey, alkaissi2023artificial}. 

As we learned from our study, BLV people watch videos for a wide range of purposes, ranging from entertainment to learning critical information that can impact their safety. Misinformation on a medical video, for example, could be life-threatening. Additionally, consider that a user might want to access a breaking news story about climate change and extreme weather events, or a video reporting on vaccine efficacy. Should their descriptions be personalized and adopt a tone similar to the user’s favorite news publication, or should they be more neutral? Should this differ for videos found on news outlets versus social media? Long-term, how would partisan audio description and video accessibility design influence people’s views, political or otherwise? 

BLV end users inevitably have varied information goals and preferences for video accessibility. As AI continues to rapidly advance, the potential for end users to have personalized agents that can learn and remember their preferences also grows. In line with its impacts on text and image generation, AI is likely to play a big role in video, and video accessibility, generation as well. We encourage future work to leverage the capabilities of generative AI, with a human in the loop, to achieve greater video accessibility at scale while mitigating potential risks and harms.

\subsection{Limitations and Future Work}
In this study, we investigated a small set of specific scenarios based on findings from our formative survey. Our survey had a relatively small number of respondents, so our survey findings should not be generalized. Due to limitations of our survey platform, the video types and platforms listed were not randomized. We acknowledge that this could have affected survey results, which in turn could have impacted our scenarios for the interview study. While studying specific scenarios slightly limits the generalizability of our findings, we highlight variation in BLV users’ preferences across a diverse set of different scenarios and provide a foundation for future video accessibility work. Additionally, as all interview participants self-identified as blind, we did not have an opportunity to thoroughly understand how visual enhancements could improve video accessibility for people with low vision \textemdash{} future work could specifically focus on low vision people’s experiences. 

We encourage researchers to continue critically examining the ethical and societal impacts of personalized and holistic video accessibility. For example, how do BLV users’ lived experiences and cultural backgrounds influence their preferences for styles and tones? What are the implications of video accessibility personalization in terms of reinforcing echo chambers or biases, especially for news and social media content? Furthermore, given the large variety and volume of video content, it is infeasible to manually create personalized accessible videos for all possible scenarios or user preferences. Future work can explore how generative AI can be applied to AD creation in practice, understand to what degree biases can manifest in AI-generated video accessibility, including representational harms \cite{bennett2021s, glazko2023autoethnographic}, and investigate how these biases can impact BLV users’ perspectives and trust of AI systems long-term. 

Lastly, this work was conducted via virtual interviews; due to logistical and software constraints, we did not examine BLV users’ video access preferences in-the-wild. We encourage future work to investigate this area with longitudinal and in-situ studies, such as a diary study to capture insights during participants’ actual video viewing sessions, to better capture the breadth of scenarios experienced by BLV users. 

\section{Conclusion}
Through a formative survey and a semi-structured interview study, we investigated BLV users’ preferences for video accessibility across diverse video scenarios. These preferences included varied levels and types of details provided, as well as additional output modalities such as audio cues, tactile elements, and visual enhancements. We identified preferred details and output modalities for different scenarios, such as watching short-form videos on Instagram to engage with friends, how-to videos on YouTube to learn how to do something, and science fiction movies on Netflix for entertainment. To our knowledge, we are one of the first to (1) contribute empirical insights capturing the diversity of BLV users’ video accessibility preferences, (2) consider a wide range of video viewing scenarios, and (3) present a design space to guide future accessible video creation and innovation. Understanding BLV users’ accessibility preferences across viewing scenarios can help us move towards more personalized and holistic video accessibility paradigms.

\begin{acks}
We thank our anonymous reviewers for their valuable feedback. We also thank Daniel Zhu, Kelly Jiang, Sharon Heung, Danielle Montour, and Emma McDonnell for their support and input throughout this project. We are grateful to our study participants for their involvement and insights. This work was partially supported by a gift from Meta (Meta Platforms, Inc.).
\end{acks}


\bibliographystyle{ACM-Reference-Format}
\bibliography{references}

\newpage
\onecolumn
\appendix
\section{All Participant-Specific Scenarios}\label{appendix:allprobes}
\begin{table*}[hbt!]
    \renewcommand{\arraystretch}{1.15}
    \begin{center}
    \caption{All 17 of our participant-specific video probes. Titles are hyperlinked to the video on YouTube. Some videos are linked at specific timestamps to capture the video start time used during our studies.}
    \Description{17 unique user-grounded video probes. Columns from left to right: Pseudonym, Type, Platform, Goal, Description, and Title / Link.}
    \label{table:allprobes}
        \begin{tabular}{>{\raggedright}p{0.1\textwidth} | >{\raggedright}p{0.105\textwidth} >{\raggedright}p{0.075\textwidth} >{\raggedright}p{0.15\textwidth} >{\raggedright}p{0.205\textwidth} >{\raggedright\arraybackslash}p{0.245\textwidth}}
        \toprule
        \textbf{Pseudonym} & \textbf{Type} & \textbf{Platform} & \textbf{Goal} & \textbf{Description} & \textbf{Title / Link} \\
        \midrule[\heavyrulewidth]
        Alice & Vlog & TVS & Engage with friends / entertainment & A day in the life of a software engineer & \href{https://www.youtube.com/watch?v=EQSNnARq-yI\&t=60s}{day in my life as a software engineer in NYC * in-office edition *} \\
        \rowcolor{gray!20}
        Blair & Sports & TVS & Entertainment & A highlight reel of a professional women’s tennis match & \href{https://www.youtube.com/watch?v=a24_VhF7814}{Beatriz Haddad Maia vs. Leylah Fernandez | 2023 Montreal Round 2 | WTA Match Highlights} \\
        Colin & Music video & TVS & Entertainment & Music video for \textit{Take On Me} by a-ha & \href{https://www.youtube.com/watch?v=djV11Xbc914}{a-ha - Take On Me (Official Video) [Remastered in 4K]} \\
        \rowcolor{gray!20}
        Diana & Video game / how-to & TVS & Learn how to do something & A playthrough of the video game, \textit{The Last of Us} & \href{https://www.youtube.com/watch?v=8tc4Uaan9Uk\&t=1897s}{What Have I Gotten Myself Into... * The Last of Us First Playthrough * Part 1} \\
        \rowcolor{gray!20}
        & Comedic & SNS & Entertainment & A short video of a dog & \href{https://www.youtube.com/shorts/jBEEsLeKoUM}{Dogs funny reaction to entering optical illusion rug! \#shorts} \\
        Emily & How-to / DIY & TVS & Learn how to do something & Tutorial for DIY desk upgrades & \href{https://www.youtube.com/watch?v=WBwHOyorIlk}{DIY Desk Upgrades} \\
        \rowcolor{gray!20}
        Felix & Action / foreign language & SS & Entertainment & A fight scene from a Korean drama, \textit{Vincenzo} & \href{https://www.youtube.com/watch?v=xySN7erAUgo}{Vincenzo Cassano -- Tailor Fight Scene} \\
        Grace & Cooking / how-to & TVS & Learn how to do something & A video explaining how to cook four different meals & \href{https://www.youtube.com/watch?v=URdX9rFIbcc}{4 Meals Anyone Can Make} \\
        & Informational & TVS & Learn about a person, event, or idea & A video about Manhattan's grid plan & \href{https://www.youtube.com/watch?v=QaIOfgz8FVY}{Where Manhattan's grid plan came from} \\
        \rowcolor{gray!20}
        Haley & Music video & TVS & Entertainment & A montage of animated characters engaging in adventure & \href{https://www.youtube.com/watch?v=BUi9vMqXxDo}{Theme Song | Elena of Avalor | @disneyjunior} \\
        Isaac & Music video & TVS & Entertainment & Music video for \textit{Bad Blood} by Taylor Swift ft. Kendrick Lamar & \href{https://www.youtube.com/watch?v=QcIy9NiNbmo}{Taylor Swift - Bad Blood ft. Kendrick Lamar} \\
        \rowcolor{gray!20}
        Julia & Live theater & TVS & Entertainment & A musical number from the Broadway show, \textit{Moulin Rouge} & \href{https://www.youtube.com/watch?v=HTbPslvdYa8\&t=55s}{Moulin Rouge! The Musical on Good Morning America} \\
        Karla & Exercise / how-to & TVS & Learn how to do something & A fitness instructor goes through a warm-up routine & \href{https://www.youtube.com/watch?v=Dku5R496Ino}{Pumped Up Cardio Warmup! (Easy, fun, at home workout)} \\
        \rowcolor{gray!20}
        Layne & Video game / how-to & TVS & Learn how to do something & A playthrough of two minigames in \textit{Mario Party} & \href{https://www.youtube.com/watch?v=FbrqhTpmM_Q}{Mario Party Superstars ALL MINIGAMES!!} \\
        Mason & Cooking / how-to & TVS & Learn how to do something & An instructional video of five minute meals & \href{https://www.youtube.com/watch?v=9_5wHw6l11o}{7 Recipes You Can Make In 5 Minutes} \\
        \rowcolor{gray!20}
        Nicki & Music video & TVS & Entertainment & Music video for \textit{You Belong With Me} by Taylor Swift & \href{https://www.youtube.com/watch?v=VuNIsY6JdUw}{Taylor Swift - You Belong With Me} \\
        Oscar & Action & SS & Entertainment & A fight scene from \textit{The Avengers} & \href{https://www.youtube.com/watch?v=SLD9xzJ4oeU}{Thor vs Hulk - Fight Scene - The Avengers (2012) Movie Clip HD} \\
        \bottomrule
        \end{tabular}
    \end{center}
\end{table*}
\newpage
\section{Summary of Findings}\label{appendix:summary}
\begin{table*}[h!]
    \renewcommand{\arraystretch}{1.15}
    \begin{center}
    \caption{Summary of desired details and output modalities for nine different scenarios, represented here by their video types. \\
    Fantastical = Science Fiction, Fantasy, and Animation \\
    Drama = Historical, Romance, Reality, and Drama
    }
    \Description{Summary of desired details and output modalities for nine different scenarios, represented here by their video types. Desired details include actions, text on screen, subjects, settings, clothing, visual aids, scene changes, facial expressions, visual effects, and equipment. Output modalities include ambient sound, additional resources, audio cues, tactile graphics, text to Braille, 3D models, background music, and vibration.}
    \label{table:summary}
        \begin{tabular}{>{\cellcolor{white}}p{0.02\textwidth} | p{0.18\textwidth} | p{1pt} c c c c c c c c c c p{1pt} }
            \toprule
             & & & How-To & Info / Edu & Short-Form & Music & Live & Personal & Fantastical & Comedy & Drama & \\
            \midrule[\heavyrulewidth]
             & Actions & & \ding{51} & \ding{51} & \ding{51} & \ding{51} & \ding{51} & \ding{51} & \ding{51} & \ding{51} & \ding{51} & \\
            \rowcolor{gray!20}
             & Text on Screen & & \ding{51} & \ding{51} & \ding{51} & \ding{51} & \ding{51} & \ding{51} & \ding{51} & \ding{51} & \ding{51} & \\
             & Subjects & & & \ding{51} & \ding{51} & \ding{51} & & \ding{51} & \ding{51} & \ding{51} & \ding{51} & \\
            \rowcolor{gray!20}
             & Settings & & & \ding{51} & \ding{51} & \ding{51} & & \ding{51} & & & & \\
             & Clothing & & & & \ding{51} & \ding{51} & & \ding{51} & \ding{51} & \ding{51} & \ding{51} & \\
            \rowcolor{gray!20}
             & Visual Aids & & \ding{51} & \ding{51} & & & \ding{51} & & & & & \\
             & Scene Changes & & \ding{51} & \ding{51} & & & \ding{51} & & & & & \\
            \rowcolor{gray!20}
             & Facial Expressions & & & & & & & & \ding{51} & \ding{51} & \ding{51} & \\
             & Visual Effects & & & & \ding{51} & \ding{51} & & & \ding{51} & & & \\
            \rowcolor{gray!20}
            \multirow{-10}{*}{\parbox[t]{2mm}{\rotatebox[origin=c]{90}{{Desired Details}}}} & Equipment & & \ding{51} & & & & & & & & & \\
             
            \midrule
            
             & Ambient Sound & & \ding{51} & \ding{51} & \ding{51} & \ding{51} & \ding{51} & \ding{51} & \ding{51} & \ding{51} & \ding{51} & \\
            \rowcolor{gray!20}
             & Additional Resources & & \ding{51} & \ding{51} & \ding{51} & \ding{51} & & & \ding{51} & & \ding{51} & \\
             & Audio Cues & & \ding{51} & \ding{51} & \ding{51} & & \ding{51} & & & \ding{51} & & \\
            \rowcolor{gray!20}
             & Tactile Graphics & & \ding{51} & \ding{51} & \ding{51} & & & & & & & \\
             & Text to Braille & & \ding{51} & & & \ding{51} & & & & & & \\
            \rowcolor{gray!20} 
             & 3D Models & & & \ding{51} & & & & & \ding{51} & & & \\
             & Background Music & & & & \ding{51} & \ding{51} & & & & & & \\
            \rowcolor{gray!20} 
            \multirow{-8}{*}{\parbox[t]{2mm}{\rotatebox[origin=c]{90}{{Output Modalities}}}} & Vibration & & & & & \ding{51} & & & & & & \\
            \bottomrule
        \end{tabular}
    \end{center}
\end{table*}

\end{document}